\documentclass[conference]{IEEEtran}
\ifCLASSINFOpdf
\else
\fi
\usepackage{cite}
\usepackage{amsmath,amssymb,amsfonts}
\usepackage{algorithmic}
\usepackage{graphicx}
\usepackage{textcomp}
\usepackage{xcolor}
\usepackage{diagbox}
\usepackage{subcaption}
\usepackage[capitalise]{cleveref}

\usepackage{tikz}
\usetikzlibrary{shapes,arrows}
\usetikzlibrary{calc}
\usepackage{bm,times, mathtools}
\DeclareMathOperator*{\argmin}{argmin}

\def\BibTeX{{\rm B\kern-.05em{\sc i\kern-.025em b}\kern-.08em
    T\kern-.1667em\lower.7ex\hbox{E}\kern-.125emX}}

\IEEEoverridecommandlockouts
\IEEEpubid{\makebox[\columnwidth]{978-1-6654-3597-0/21/\$31.00~\copyright2021 IEEE \hfill}\hspace{\columnsep}\makebox[\columnwidth]{}}
\begin{document}
%
\author{\IEEEauthorblockN{Jochen Stiasny, George S. Misyris, Spyros Chatzivasileiadis}
\IEEEauthorblockA{Department of Electrical Engineering\\
Technical University of Denmark, Kgs. Lyngby, Denmark \\
\{jbest, gmisy, spchatz\}@elektro.dtu.dk}}

\title{Physics-Informed Neural Networks for Non-linear System Identification for Power System Dynamics\\
\thanks{This work is supported by the multiDC project funded by Innovation Fund Denmark, Grant No. 6154-00020B.}}


%


\maketitle
\IEEEpubidadjcol

\begin{abstract}
Varying power-infeed from converter-based generation units introduces great uncertainty on system parameters such as inertia and damping. As a consequence, system operators face increasing challenges in performing dynamic security assessment and taking real-time control actions. Exploiting the widespread deployment of phasor measurement units (PMUs) and aiming at developing a fast dynamic state and parameter estimation tool, this paper investigates the performance of Physics-Informed Neural Networks (PINN) for discovering the frequency dynamics of future power systems. PINNs have the potential to address challenges such as the stronger non-linearities of low-inertia systems, increased measurement noise, and limited availability of data. The estimator is demonstrated in several test cases using a 4-bus system, and compared with state of the art algorithms, such as the Unscented Kalman Filter (UKF), to assess its performance.
\end{abstract}

\begin{IEEEkeywords}
Physics-informed neural networks, system identification, state estimation, swing equation
\end{IEEEkeywords}

%

\section{Introduction}
Dynamic security assessment of power systems heavily relies on the accuracy of system models, as well as the knowledge of system parameters that play a crucial role on the system stability \cite{kundur2004definition}. However, the gradual replacement of synchronous generators with converter-connected devices of varying generation and demand (e.g. solar PV, wind turbines, electric vehicles) leads to (i) faster frequency dynamics with larger frequency deviations, and (ii) substantially higher uncertainty in system parameters, such as inertia and damping, that become difficult to track \cite{milano2018foundations,Misyris2018}. 

Dynamic state estimation in power systems has been designed primarily around the Bayesian approach of an a priori probability distribution function for the system states that are subsequently updated with the likelihood of a measurement to form an a posteriori probability distribution \cite{Zhao2019}. This leads to the family of filtering methods, such as the Kalman filter and its derivatives, e.g. the unscented Kalman filter. For parameter estimation, the filter not only includes the internal states of the system but is usually augmented with the unknown system parameters. These filter-based methods offer the potential for execution in real-time which allows for online parameter estimation and state tracking, as well as fault detection. However, the handling of strong non-linearity, and the challenging initialisation constrain the applicability of these methods. 

To overcome the challenges of these methods, the use of feed-forward neural networks, primarily for static state estimation, has been proposed in the literature \cite{Mestav_PESGM2018, Barbeiro_FNN_SE}. Although neural network training procedures coupled with some physical models have been applied for the static state estimation of distribution systems \cite{Giannakis_PIDNN}, the approaches are overwhelmingly model agnostic. In particular when it comes to dynamic state and parameter estimation of time-varying systems, conventional neural networks do not address two main challenges: First, the underlying parameters are unknown. As a result, we are dealing by design with an unsupervised learning problem, as there does not exist a known desired outcome to train against. Second, reliable, structured, and extensive data sets are often not available. These two fundamental obstacles have been recently addressed by physics-informed neural networks (PINNs), which can incorporate the underlying physical laws that the dynamic system shall comply with in the training procedure \cite{RAISSI2019686}. In our previous work, we have developed and applied physics-informed neural networks for the dynamic model of a single machine infinite-bus system with the focus on approximating the swing equation's solution \cite{Misyris2019}.

\section{System model \& Methodology} \label{Meth}
In this section we introduce the dynamic system and how the PINN is built up and trained.

\subsection{Dynamic System Model}

We consider a dynamic model for a power system that can be expressed in the form of a Differential Algebraic Equation (DAE) system: 
\begin{align}
    \centering
    \dot{\mathbf{x}} &= \mathbf{f}(\mathbf{x}, \mathbf{y}, \mathbf{r}; \boldsymbol{\lambda}) \label{eq::nonlinear_function} \\
    \boldsymbol{0} &= \mathbf{g}(\mathbf{x}, \mathbf{y}, \mathbf{r}; \boldsymbol{\lambda})
\end{align}
where $\mathbf{f}(\mathbf{x}, \mathbf{y}, \mathbf{r}; \boldsymbol{\lambda})$ corresponds to the first order non-linear differential equations of the system, and $\mathbf{g}(\mathbf{x}, \mathbf{y}, \mathbf{r}; \boldsymbol{\lambda})$ to the algebraic equations relating the outputs of the system to its state variables and inputs. Vector $\mathbf{x}$ represents the dynamic states, $\mathbf{y}$ the algebraic variables, $\mathbf{r}$ the input variables, and $\boldsymbol{\lambda}$ are the system parameters. 

In this paper, we focus on the power system frequency dynamics, which in their simplest and most common form, are described by the swing equation. For each generator $k$, the resulting system of equations can then be represented by \cite{PESGM_2016A,Misyris2018}:
\begin{equation}
\centering
    m_k\ddot{\delta}_k+d_k\dot{\delta}_k+\sum_jB_{kj}V_kV_j\sin(\delta_k-\delta_j)-P_k=0
    \label{Eq::SwingEquation}
\end{equation}
where $m_k$ defines the generator inertia constant, $d_k$ represents the damping coefficient, $B_{kj}$ is the $\{k,j\}\textnormal{-entry}$ of the bus susceptance matrix, $P_k$ is the mechanical power of the $k^{\rm th}$ generator, $V_k, V_j$ are the voltage magnitudes at buses $k,j$ and $\delta_k, \delta_j$ represent the voltage angles behind the transient reactance, for generators this is the rotor angle. $\dot\delta_k$ is the angular frequency of generator $k$, often also denoted as $\omega_k$. 

For frequency dependent loads, \eqref{Eq::SwingEquation} simplifies to:
\begin{equation}
\centering
    d_k\dot{\delta}_k+\sum_jB_{kj}V_kV_j\sin(\delta_k-\delta_j)-P_k=0
    \label{Eq::SwingEquation_load}
\end{equation}
where usually $P_k < 0$. For brevity, the term $B_{kj}V_kV_j$ will be referred to as connectivity $a_{ij}$.

In this paper, we seek to determine uncertain parameters, $m_k$ and $d_k$, and to estimate the rotor angles $\delta_k$ and generator speeds $\omega_k$. For this we use a set of measurements capturing the temporal evolution of a power system. These could be earlier recorded data or currently observed dynamics.

\subsection{Neural networks as function approximators} \label{subsec:state_predictor}

A neural network $\mathbf{u}(t)$ essentially approximates a function, in this case, the temporal state evolution $\mathbf{h}(t)$ of the dynamic system which is the solution to the system of DAEs given an initial condition $\mathbf{x}_0$.
\begin{align}
    \mathbf{u}(t) \approx \mathbf{h}(t; \mathbf{x}_0, \mathbf{\lambda}) = \delta_i (t; \mathbf{x}_0, \mathbf{\lambda}).
\end{align}
By supplying data points we adjust the parameters the neural network to achieve a good fit. This results in a continuous approximation function $\mathbf{u}(t)$, however, for time instances in between the given data points the neural network interpolates in a model agnostic, hence non-physical way.
The network to represent $\mathbf{u}(t)$ consists of $N_L$ fully-connected hidden layers with $N^{i}$ neurons in the $i^{\rm th}$ layer. In each hidden layer we apply the operation
\begin{equation}\label{eq::hidden_layer}
    \begin{aligned}
        \mathbf{y}^i = &\; \sigma \left(\mathbf{W}^i \mathbf{y}^{i-1} + \mathbf{b}^{i}\right),\; i \in \{1, ..., N_L\},
    \end{aligned}
\end{equation}
where $\mathbf{y}^i$ represents a hidden layer's output, $\mathbf{y}^{i-1}$ the hidden layer's input, $\mathbf{W}^i$ a weight matrix and $\mathbf{b}^i$ a bias vector. The activation function $\sigma$ introduces non-linearities as we choose the hyperbolic tangent. The input to the network $\mathbf{y}^0$ is simply time $t$. For the output $\mathbf{u}$ we once more apply \eqref{eq::hidden_layer}, however, here $\sigma$ is chosen to be the identity function.
By using automatic differentiation (AD) we can obtain derivatives of $\mathbf{u}(t)$ with respect to the input $t$, i.e. $\dot{\mathbf{u}} \coloneqq \frac{\partial}{\partial t}\mathbf{u}(t)$ which should resemble $\omega_k(t)$. Consequently, we fit the network by comparing the approximations of $\mathbf{u}$ and $\dot{\mathbf{u}}$ with measurements for $\delta_k$ and $\omega_k$, represented as $z_k^n$ and $\dot{z}_k^n$, at $N_z$ given time instances. This forms the loss $\mathcal{L}_z$
\begin{equation}\label{eq::measurement_loss}
    \centering
    \mathcal{L}_z \coloneqq \frac{1}{N_z}\sum_{n=1}^{N_z} \sum_{k=1}^{N_{k}} \left(z_k^n - u_k(t_n) \right)^2 + \left(\dot{z}_k^n - \dot{u}_k(t_n) \right)^2
\end{equation}
which shall be minimised by adjusting the networks parameters $\mathbf{W}^i$ and $\mathbf{b}^i$.

\begin{figure}[tbp]
    \centering
     \scalebox{0.75}{\pagestyle{empty}
\def\layersep{1.5cm}
\def\nodeinlayersep{1cm}
\begin{tikzpicture}[
   shorten >=1pt,->,
   draw=black!70,
    node distance=\layersep,
    every pin edge/.style={<-,shorten <=1pt},
    neuron/.style={circle,fill=black!25,minimum size=17pt,inner sep=0pt},
    input neuron/.style={neuron, fill=blue!30, minimum size=25pt,draw=black},
    output neuron/.style={neuron, fill=red!50, minimum size=25pt,draw=black},
    hidden neuron/.style={neuron, fill=gray!30},
    operator neuron/.style={neuron, fill=yellow!50, minimum size=25pt,draw=black},
    summation neuron/.style={neuron, fill=gray!50, minimum size=18pt},
    parameter neuron/.style={neuron, fill=green!30, minimum size=25pt,draw=black},
    annot/.style={text width=4em, text centered}
]
    
    \node[input neuron] (I-1) at (0,-3*\nodeinlayersep) {t};  

    \node[hidden neuron] (H1-1) at (1*\layersep,-1*\nodeinlayersep ) {$\sigma$};
    \node[hidden neuron] (H1-2) at (1*\layersep,-2*\nodeinlayersep ) {$\sigma$};
    \node[hidden neuron] (H1-3) at (1*\layersep,-3*\nodeinlayersep ) {$\sigma$};
    \node (H1-4) at (1*\layersep,-4*\nodeinlayersep ) {$\vdots$};
    \node[hidden neuron] (H1-5) at (1*\layersep,-5*\nodeinlayersep ) {$\sigma$};
    
    \node[hidden neuron] (H2-1) at (2*\layersep,-1*\nodeinlayersep ) {$\sigma$};
    \node[hidden neuron] (H2-2) at (2*\layersep,-2*\nodeinlayersep ) {$\sigma$};
    \node[hidden neuron] (H2-3) at (2*\layersep,-3*\nodeinlayersep ) {$\sigma$};
    \node (H2-4) at (2*\layersep,-4*\nodeinlayersep ) {$\vdots$};
    \node[hidden neuron] (H2-5) at (2*\layersep,-5*\nodeinlayersep ) {$\sigma$};
    
    \node[output neuron] (O-1) at (3*\layersep,-1.2*\nodeinlayersep) {$u_1$};
    \node[output neuron] (O-2) at (3*\layersep,-2.4*\nodeinlayersep) {$u_2$};
    \node (O-3) at (3*\layersep,-3.6*\nodeinlayersep) {$\vdots$};
    \node[output neuron] (O-4) at (3*\layersep,-4.8*\nodeinlayersep) {$u_k$};

    \node[operator neuron] (D-1) at (4.2*\layersep,-1.5*\nodeinlayersep) {$I$};
    \node[operator neuron] (D-2) at (4.2*\layersep,-3*\nodeinlayersep) {$\frac{\partial}{\partial t}$};
    \node[operator neuron] (D-3) at (4.2*\layersep,-4.5*\nodeinlayersep) {$\frac{\partial^2}{\partial t^2}$};
    
   \node [rounded corners=0.2cm] (measurementLoss) at (5.3*\layersep,-1.5*\nodeinlayersep) [draw,fill=orange!30,thick,minimum width=1cm,minimum height=1cm] {$\mathcal{L}_z$};
   \node [rounded corners=0.2cm] (physicsLoss) at (5.3*\layersep,-4.5*\nodeinlayersep) [draw,fill=orange!30,thick,minimum width=1cm,minimum height=1cm] {$\mathcal{L}_c$};
   
   \node [summation neuron] (totalLoss) at (5.3*\layersep,-3*\nodeinlayersep) {$+$};
   
   \node [parameter neuron] (optimal_variable) at (6.3*\layersep,-3*\nodeinlayersep) {$\hat{\lambda}^*$};

    \foreach \source in {1}
        \foreach \dest in {1,...,3,5} 
            \path (I-\source) edge (H1-\dest);
    
    \foreach \source in {1,...,3,5}
        \foreach \dest in {1,...,3,5} 
            \path (H1-\source) edge (H2-\dest);
            
    \foreach \source in {1,...,3,5}
        \foreach \dest in {1,2,4} 
            \path (H2-\source) edge (O-\dest);
            
    \foreach \source in {1,2,4} 
        \foreach \dest in {1,2,3}
            \path (O-\source) edge (D-\dest);
            
    \path (D-1) edge (measurementLoss);
    \path (D-2) edge (measurementLoss);
    \path (D-1) edge (physicsLoss);
    \path (D-2) edge (physicsLoss);
    \path (D-3) edge (physicsLoss);
    
    \draw (measurementLoss.south)--(totalLoss.north);
    \draw (physicsLoss.north)--(totalLoss.south);
    
    \draw (totalLoss.east) -- (optimal_variable) node [midway, above=4pt, fill=white] {min};
    \node [rounded corners=0.2cm] (NeuralNetwork) at (1.5*\layersep,-3*\nodeinlayersep ) [draw,black, dashed, very thick,minimum width=4*\layersep,minimum height=5*\nodeinlayersep] {};
    \node [rounded corners=0.2cm] (AD) at (D-2) [draw,black, dashed, very thick,minimum width=1*\layersep,minimum height=5*\nodeinlayersep] {};
    \node[annot, above=2pt] at (NeuralNetwork.north) {NN};
    \node[annot, above=2pt] at (AD.north) {AD};
\end{tikzpicture}}  
 
    \caption{Physics-informed neural network architecture}
    \label{fig:pinn_architecture}
\end{figure}
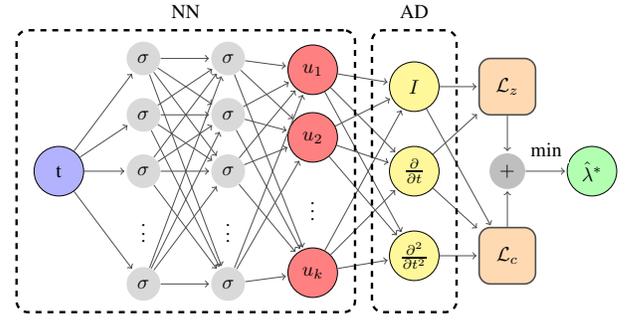

\subsection{Physics regularisation}
The function approximation presented above falls in the category of supervised learning where we train the network against a desired output: the measurements. However, when the desired output, i.e. the system parameters, is unknown, the problem becomes an unsupervised learning task. In this part of the training, we aim to find parameter estimates $\hat{m}_k$ and $\hat{d}_k$ that yield the best agreement of the underlying physical equations \eqref{Eq::SwingEquation} and \eqref{Eq::SwingEquation_load} when checked against the NN output $u_k(t)$ and its temporal derivatives $\dot{u}_k(t)$ and $\ddot{u}_k(t)$. We express this notion in the `consistency' $f_k(t)$ at each bus, effectively applying \eqref{Eq::SwingEquation} to the NN outputs:
\begin{equation}    
\centering
\begin{aligned}
    f_k(t) \coloneqq \; & \hat{m}_k \ddot{u}_k(t)+\hat{d}_k \dot{u}_k(t) \\
    & + \sum_j a_{kj} \sin(u_k(t) - u_j(t))-P_k
\end{aligned}
\label{Eq::physics_consistency}
\end{equation}
Our aim is to achieve $f_k(t) =  0$, meaning that the NN output will comply with the power system dynamics described by the swing equation. Hence, minimising the regularisation term $\mathcal{L}_c$ during training shall lead to neural networks that obey the underlying physical laws of power systems, i.e. physics-informed neural networks:
\begin{equation}\label{eq::collocation_loss}
    \centering
    \mathcal{L}_c \coloneqq \frac{1}{N_c}\sum_{n=1}^{N_c} \sum_{k=1}^{N_{k}} f_k(t_c)^2
\end{equation}
It is important to note that the evaluation of this loss function is not bound to the measured data. Instead it can be evaluated for any time input. We do so by generating $N_c$ equally spaced points across the time domain; we refer to them as collocation points. Furthermore, the value of $\mathcal{L}_c$ can be reduced either by better parameter estimates $\hat{m}_k$ and $\hat{d}_k$ or by a more physically `consistent' approximation of $u_k(t),\dot{u}_k(t)$ and $\ddot{u}_k(t)$. Hence, $\mathcal{L}_c$ also affects the weight matrices and bias vectors of the NN. Due to this influence on the NN's parameters, $\mathcal{L}_c$ can be regarded as a regularisation term which biases the approximation towards a more physically `consistent' one. 

\subsection{Training process}

By minimising the total loss function $\mathcal{L}_z + \mathcal{L}_c$ over the variables $\mathbf{W}^i, \mathbf{b}^i, \hat{m}_k$ and $\hat{d}_k$, we obtain estimates $\hat{m}_k^*$ and $\hat{d}_k^*$ for the underlying parameters:
\begin{equation}\label{eq:total_loss}
    \hat{m}_k^*, \hat{d}_k^* = \argmin_{\mathbf{W}^i, \mathbf{b}^i, \hat{m}_k, \hat{d}_k} \mathcal{L}_z + \mathcal{L}_c
\end{equation}
However, in practice, \eqref{eq:total_loss} poses a highly non-convex and multi-parameter optimisation problem as usually encountered in NN training. We use the Adam optimiser \cite{kingma2017adam}, a first-order stochastic gradient-based method, to obtain a solution. The algorithm involves stepwise adjustments of the optimisation variables according to the computed gradient until a sufficiently stable solution is obtained. This requires multiple `epochs', which is simply the number of times the entirety of the available data (collocation points and measurements) has been used in the optimisation. To speed up or even make the training feasible we apply batching of the data points. The batch size is varied during the training, starting with small batches and ending with an optimisation over the whole data set. In practice, this results in initially large adjustments of the parameters to identify the valley of the objective function around the true values such that the final estimation of the parameters can be accurate and is not trapped in another local minimum.

\section{Simulation}\label{Sim}
In this section, we briefly introduce the used power system, the training procedure and the unscented Kalman filter, that serves as a comparison.

\subsection{4-Bus 2-Generator Model}
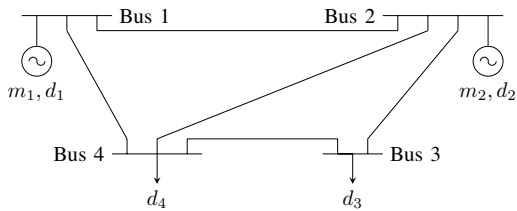
\begin{figure}[htb]
    \centering
     \scalebox{0.8}{\tikzset{sin v source/.style={
  circle,
  draw,
  append after command={
    \pgfextra{
    \draw
      ($(\tikzlastnode.center)!0.5!(\tikzlastnode.west)$)
       arc[start angle=180,end angle=0,radius=0.425ex] 
      (\tikzlastnode.center)
       arc[start angle=180,end angle=360,radius=0.425ex]
      ($(\tikzlastnode.center)!0.5!(\tikzlastnode.east)$) 
    ;
    }
  },
  scale=1.5,
 }
}

\begin{tikzpicture}
\draw
(0,0) node [sin v source] (v1) {} 
(v1.north)--++(0,0.5)  
coordinate(v1-gen)
(v1-gen)--++(-0.25,0) 
(v1-gen)--++(0.5, 0)
coordinate(v1-v4)
(v1-v4)--++(0, -0.25)
coordinate(v1-v4-offset)
(v1-v4)--++(0.5, 0)
coordinate(v1-v2)
(v1-v2)--++(0.25,0) node[right]{Bus 1} 
(v1-v2)--++(0, -0.25)
coordinate(v1-v2-offset)

(v1-v2-offset)--++(5,0)
coordinate(v2-v1-offset)
(v2-v1-offset)--++(0,0.25)
coordinate(v2-v1)
(v2-v1)--++(-0.25,0) node[left]{Bus 2}
(v2-v1)--++(0.5,0)
coordinate(v2-v4)
(v2-v4)--++(0, -0.25)
coordinate(v2-v4-offset)
(v2-v4)--++(0.5,0)
coordinate(v2-v3)
(v2-v3)--++(0, -0.25)
coordinate(v2-v3-offset)
(v2-v3)--++(0.5,0)
coordinate(v2-gen)
(v2-gen)--++(0, -0.5) node[below,sin v source]{}
(v2-gen)--++(0.25, 0)

(v1-v4-offset)--++(1, -1.8)
coordinate(v4-v1-offset)
(v4-v1-offset)--++(0, -0.25)
coordinate(v4-v1)
(v4-v1)--++(-0.25,0) node[left]{Bus 4}
(v4-v1)--++(0.5, 0)
coordinate(v4-v2)
(v4-v2)--++(0, 0.25)
coordinate(v4-v2-offset)
(v4-v2-offset)--(v2-v4-offset)
(v4-v2)--++(0.5, 0)
coordinate(v4-v3)
(v4-v3)--++(0, 0.25)
coordinate(v4-v3-offset)
(v4-v3)--++(0.25, 0)

(v4-v3-offset)--++(2.5,0)
coordinate(v3-v4-offset)
(v3-v4-offset)--++(0, -0.25)
coordinate(v3-v4)
(v3-v4)--++(-0.25,0)
(v3-v4)--++(0.5,0)
coordinate(v3-v2)
(v3-v2)--++(0,0.25)
coordinate(v3-v2-offset)
(v3-v2-offset)--(v2-v3-offset)
(v3-v2)--++(0.25,0) node[right]{Bus 3}
(v3-v4)--++(0.25,0)
coordinate(v3-load)
;

\draw[-stealth](v4-v2)--++(0,-0.5cm); 
\draw[-stealth](v3-load)--++(0,-0.5cm); 
\node[] at (0,-0.5) {$m_1, d_1$};
\node[] at (7.5,-0.5) {$m_2, d_2$};
\node[] at (2,-2.3) {$d_4$};
\node[] at (5.25,-2.3) {$d_3$};
\end{tikzpicture}}  
  
    \caption{4-bus 2-generator system}
    \label{fig:4_bus_system}
\end{figure}

A 4-bus system with two generators as depicted in \cref{fig:4_bus_system} is used as a test case. The small system size shall facilitate the understanding as we can highlight the characteristics of PINNs. We test three parameter sets, denoted as systems A, B, and C, and shown in \cref{tab:4_bus_parameters}. The three variants shall illustrate how the investigated methods perform in different setups, namely a `standard' system A, a system with faster dynamics (system B), and a system with slower dynamics (system C) compared to the measurement frequency.

\begin{table}[tb]
\centering
\caption{Parameters of the 4-bus 2-generator test case}
\label{tab:4_bus_parameters}
\begin{tabular}{|l|lll|}
\hline
System Parameters & A & B & C \\ \hline
$m_1$ (p.u.) & 0.3 & 0.02 & 5.2 \\
$m_2$ (p.u.) & 0.2 & 0.03 & 4.0 \\
$d_1$ (p.u.) & 0.15 & 0.01 & 3.8 \\
$d_2$ (p.u.) & 0.3 & 0.015 & 4.3 \\
$d_3$ (p.u.) & 0.25 & 0.02 & 10.5 \\
$d_4$ (p.u.) & 0.25 & 0.04 & 8.3 \\
$a_{13}$ (p.u.) & 0.5 & 0.5 & 2.5 \\
$a_{14}$ (p.u.) & 1.2 & 1.2 & 2.2 \\
$a_{23}$ (p.u.) & 1.4 & 1.0 & 2.0 \\
$a_{24}$ (p.u.) & 0.8 & 0.8 & 4.8 \\
$a_{34}$ (p.u.) & 0.1 & 0.1 & 0.7 \\ \hline
\end{tabular}
\end{table}

Furthermore, we assume to start from an unperturbed state, i.e. $\mathbf{\dot{x}} = 0$ and then perturb it by the constant input signal $P_k = \{0.1, 0.2, -0.1, -0.2\} p.u.$ for $t > 0$.

\subsection{Unscented Kalman Filter}
Various nonlinear filters have been used with the Kalman filter framework and applied to power systems. Among them, the Unscented Kalman Filter (UKF) has been extensively employed for parameter estimation in power systems \cite{Zhao2019} and therefore shall serve as a comparison for our method. The UKF is centred around the idea to describe the state $\mathbf{x}$ as a distribution defined by a small number of characteristic points. These so-called sigma points are passed through the non-linear function \eqref{eq::nonlinear_function}, are then weighted, and finally used to form a state prediction together with a covariance and cross-correlation matrix. In the update step, the predictions are combined with the measurements to form the new estimate. A detailed description of the UKF can be found in \cite{Ariff2014}; here, we follow its notation. 
In order to estimate $m_i^k$ and $d_i^k$, we extend the system state $\mathbf{x}^k = \left[\delta_i^k, \omega_i^k \right]^T$ by including the unknown parameters $\mathbf{\psi}^k = \left[ m_i^k, d_i^k \right]^T$. The discrete state updates are formulated by discretising \eqref{Eq::SwingEquation} and \eqref{Eq::SwingEquation_load}.

\subsection{Training procedure}

The choice of the NN size plays a key role in how accurately the function approximation can capture non-linearities in the dynamic system. We chose a PINN with two hidden layers with 30 neurons, such that all systems can be accurately represented over the input time interval $t \in [0, 2]s$.

Phasor-Measurement-Units (PMUs) allow us to record the voltage angles\footnote{In this work we neglect the effect of the transient reactance, hence the voltage angle and the rotor angle of a generator align.} $\delta$ and derive from it the angular frequency $\dot{\delta}$ as specified in the IEEE/IEC International Standard \cite{PMUstandard}. The sampling frequency for the measurements can range between 10 and 100 frames per second for a 50 Hz system. We simulate measurements every 0.01 seconds, resulting in 200 measurements. We start splitting the data (including 20 collocation points per measurement) in batches of size 200 and then increase them consecutively to \{400, 800, 2000, 4000\} data points. The batches are trained for \{100, 200, 400, 1000, 4000\} epochs respectively. Since the NN's weights are initialised randomly we run each estimation 20 times. If not mentioned otherwise, these training parameters are applied to each simulation. To test the PINN robustness to noise, as we will see in the next section, we investigate the performance of the methods when we add Gaussian noise with $\sigma$ up to 5\% of the measurement's value, or a uniform noise distribution of up to $\pm 5\%$. 

\section{Results} \label{Res}
In this section we present key features of PINNs by investigating their behaviour in response to different systems, input data, and noise.

\subsection{Parameter estimation accuracy}

\begin{figure}[tb]
    \centering
    \includegraphics[width=\linewidth]{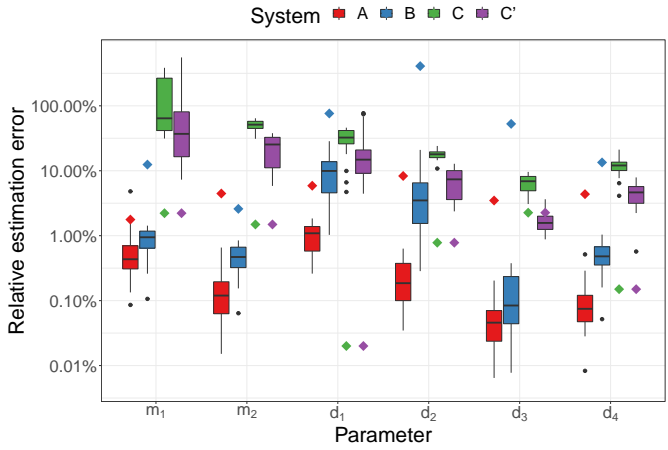}
    \caption{Accuracy of parameter estimates for PINNs (boxplots) and UKF (diamonds) for the different systems with standard (A), fast (B), and slow dynamics (C)}
    \label{fig:accuracy_parameter_estimation}
\end{figure}

The main objective is to obtain reliable and accurate estimates for the system parameters $\boldsymbol{\lambda}$. \Cref{fig:accuracy_parameter_estimation} shows that PINNs for system A as well as for some parameters in system B achieve a relative error below 1\% - performing similarly or better than a UKF. Especially for system B, which is a low-inertia system exhibiting fast dynamics, setting up a UKF proves to be challenging since the `high-information' dynamics occur within only a few time steps which the UKF struggles to capture. Although this also challenges the PINNs, they are still able to estimate a reasonable set of parameters. For slow dynamics, as seen in system C, the UKF recovers the parameters very well in contrast with PINNs. PINNs face the issue that the optimisation landscape is very flat, i.e. no direction of the gradient dominates, and so the adjustment of the weights and parameters in each epoch becomes very small. Eventually, the estimates are either trapped in a local minimum or slowly converge to the underlying parameters. Using a NN with less neurons can partially address this issue as system C' in \cref{fig:accuracy_parameter_estimation} shows. 

\begin{figure}[ht]
    \begin{subfigure}{\linewidth}
        \centering
        \includegraphics[width=\linewidth]{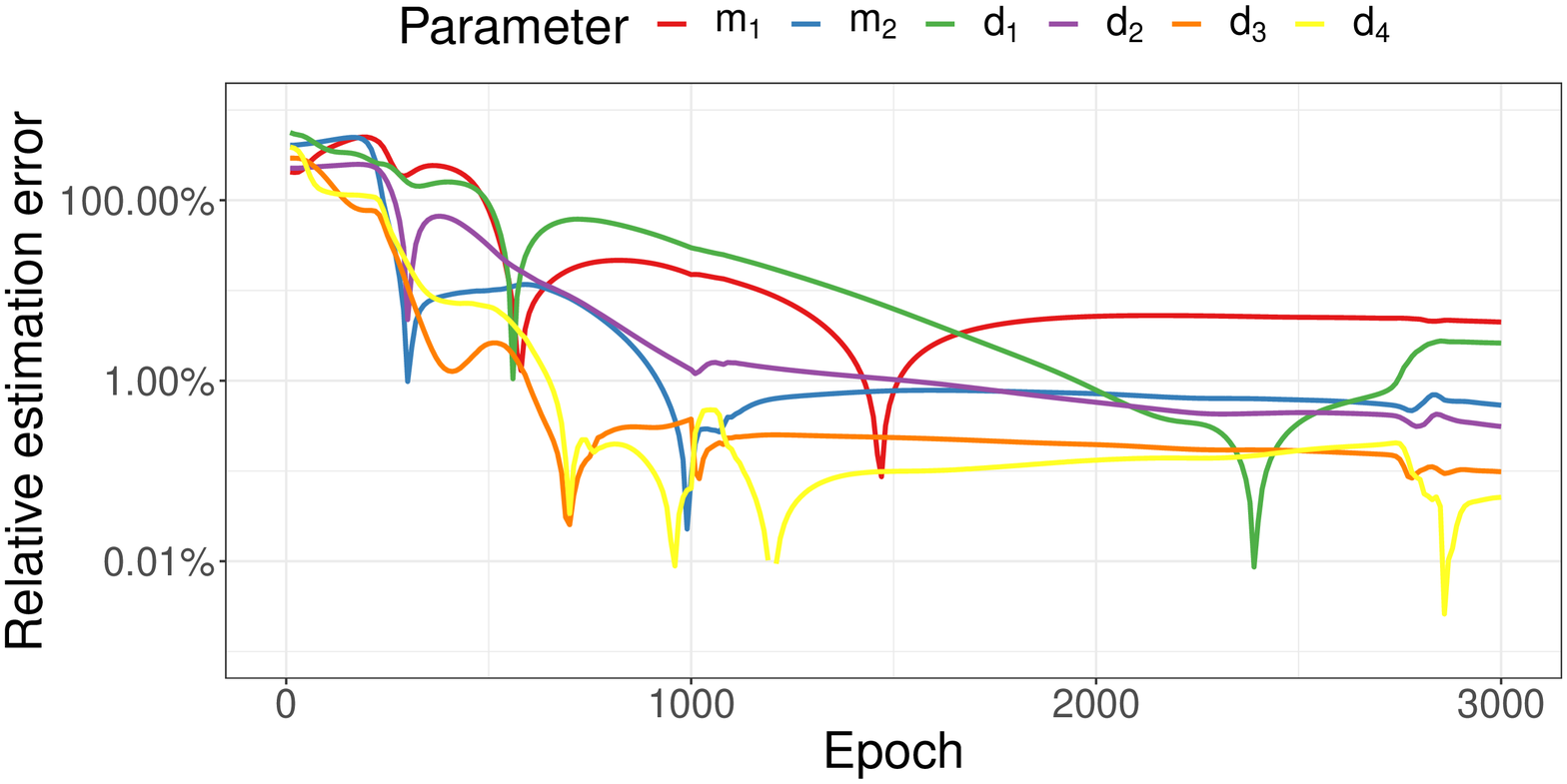}
        \caption{PINNs}
        \label{fig:evolution_parameter_pinn}
    \end{subfigure}
    
    \begin{subfigure}{\linewidth}
        \centering
        \includegraphics[width=\linewidth]{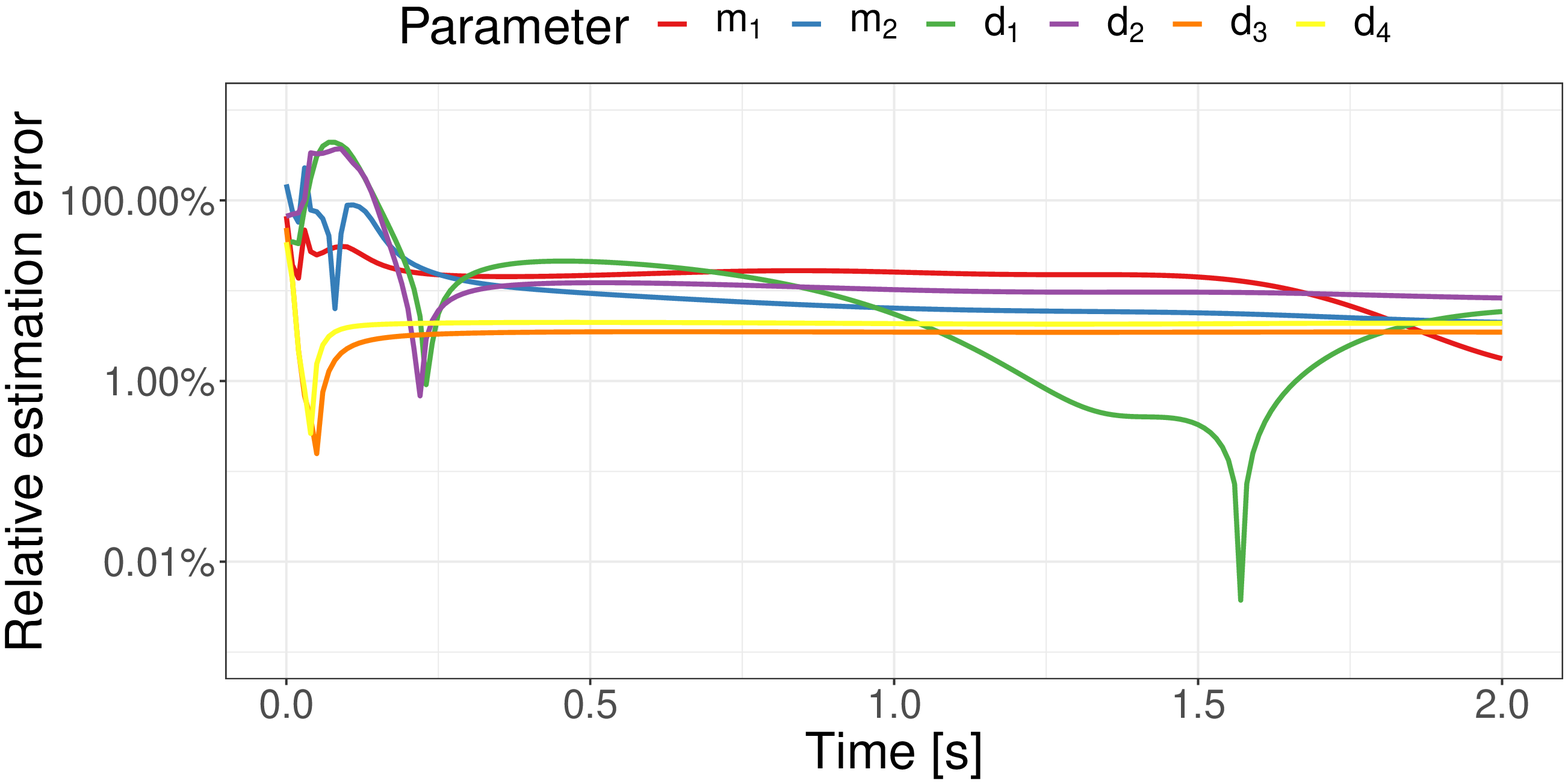}
        \caption{UKF}
        \label{fig:evolution_parameters_UKF}
    \end{subfigure}
    \caption{Evolution of parameter estimates for PINNs and UKF}
    \label{fig:evolution_parameter_estimation}
\end{figure}

The convergence of the two methods is best shown as an evolution over epochs for PINNs and an evolution over time for the UKF, as illustrated in \cref{fig:evolution_parameter_estimation}. The different x-axes hint at what distinguishes the two methods. The parameter estimates of the UKF, like other filter-based methods, evolve over time as each incoming measurement involves a new filter prediction and update of the parameter estimates. The PINN, in contrast, evolves over epochs, i.e. the number of times the complete data set is used, since each epoch involves the optimisation over the loss function and then a (small) update of all variable parameters. Hence, as long as the optimisation is not stuck at a local minimum, additional epochs should drive the estimate closer to the underlying parameters.  This repetitive use of the same data set over and over again illustrates a key difference between the two methods: the relation between data points sits at the core of PINNs in contrast to discrete filters that look at each measurement separately.  

\subsection{State approximation accuracy}

\begin{figure}[tb]
    \centering
    \includegraphics[width=\linewidth]{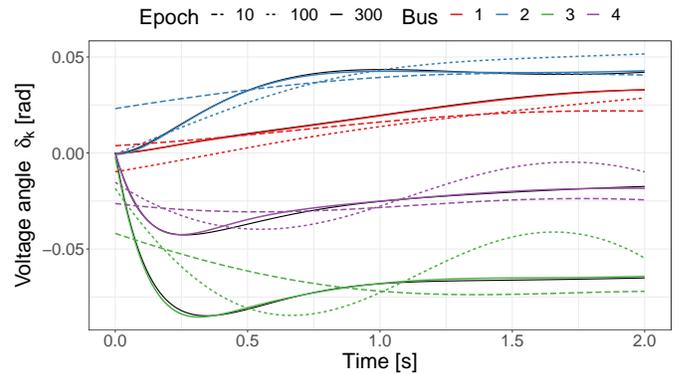}
    \caption{State estimates of PINN during the training process (black solid lines show the true trajectory).}
    \label{fig:state_estimate_evolution_pinn}
\end{figure}

As described in \cref{subsec:state_predictor}, a part of the PINN acts as a state approximation similar to the state estimates of the UKF. This approximation initially does not capture the dynamics, however, after a few epochs it matches the true system dynamics very closely as shown in \cref{fig:state_estimate_evolution_pinn}. This accurate state approximation is furthermore not dependent on an accurate parameter estimation as we can observe in \cref{fig:evolution_parameter_pinn} at the presented epochs due to the dominating loss term $\mathcal{L}_z$. In later epochs, the loss term $\mathcal{L}_c$ becomes more relevant, hence the parameter estimates improve.
A central feature of the PINN as a state approximator is that the system state can be directly extracted at any desired time instance within the time window without the need for interpolation or linearisation of system. This property stems from the continuous and differentiable nature of the PINN and is a fundamental difference compared to the discrete `predict-update' formalism of the UKF. Finally, a key observation related to the PINN's performance is that the estimator's nature prevents over-fitting with respect to measurements since the loss $\mathcal{L}_c$ effectively `smoothens' the state approximation. While this is less relevant for measurements with little noise, it becomes important for higher noise levels as it will be shown in the next section. 

\subsection{Performance under noise}
\begin{figure}[tb]
    \centering
    \includegraphics[width=\linewidth]{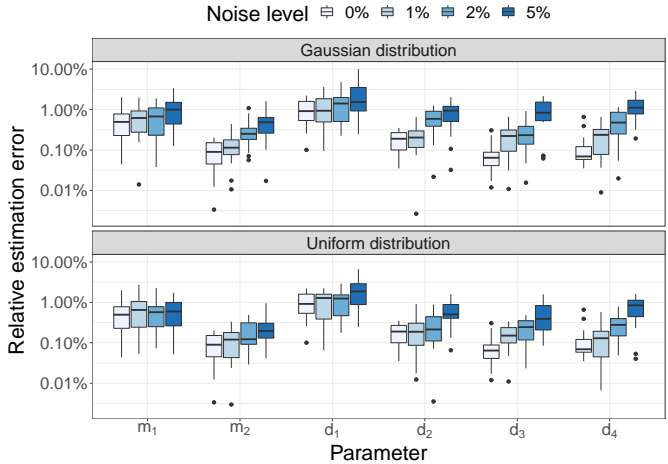}
    \caption{Parameter estimation accuracy of PINNs for different levels and types of measurement noise}
    \label{fig:noise_resistance_pinn}
\end{figure}

The robustness to noise has two aspects: how noisy are the measurements and how is the noise distributed. Regarding the noise level, as we would expect, we observe less accurate parameter estimates for increasing noise levels (\cref{fig:noise_resistance_pinn}). Nevertheless, the variation of the parameters remains small across different initialisations of the NN's weight matrices $\mathbf{W}^i$. We furthermore observe in \cref{fig:noise_resistance_pinn} that the distribution type of the noise seems to have a minor impact. These two observations link closely to the physics regularisation. It prevents potential over-fitting of the state approximation which in turn makes the parameter estimates more consistent. Meanwhile, we do not rely on any assumptions on the noise level or distribution. Conversely, by adapting the loss function $\mathcal{L}_z$, one could easily incorporate knowledge about the noise to improve the performance and robustness. This once more highlights the great flexibility compared to filter-based methods that are often specifically build around a certain assumption on the noise distribution.

\subsection{Data dependency}

\begin{figure}[tb]
    \centering
    \includegraphics[width=\linewidth]{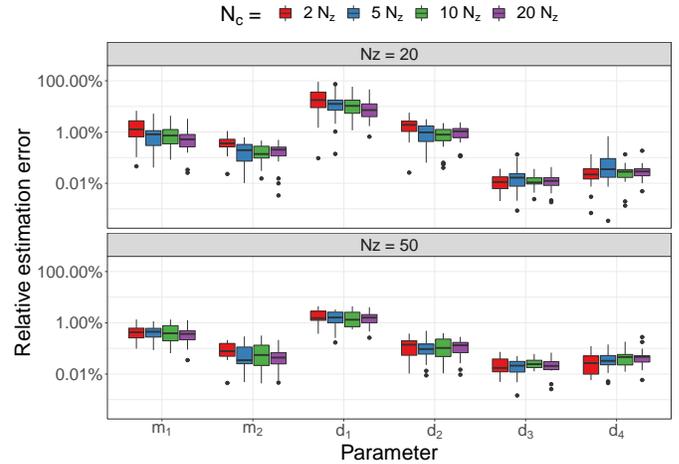}
    \caption{Parameter estimation accuracy of PINNs for different numbers of measurement and collocation points $N_z$ and $N_c$.}
    \label{fig:data_dependency_pinns}
\end{figure}

In this subsection we illustrate how PINNs behave under different data sets. First, we reduce the observed time span from 2 seconds down to 0.2 and 0.5 seconds and hence the number of measurements $N_z$ decreases by a factor of 10 and 4 respectively. A possibly counter-intuitive feature of PINNs is displayed in \cref{fig:data_dependency_pinns}, as the estimation accuracy remains high for parameters apart from $d_1$. Furthermore we can observe that the number of collocation points $N_c$ expressed as a multiple of measurement points $N_z$ in \cref{fig:data_dependency_pinns} mainly plays a role if few measurements are available. In these cases, we evaluate the consistency of our estimation with the underlying physics at more points and thereby achieve better estimates. However, this effect is strongest for the first collocation points and then levels off. This is important in the training process since more data points require more time per epoch.

\begin{figure}[tb]
    \centering
    \includegraphics[width=\linewidth]{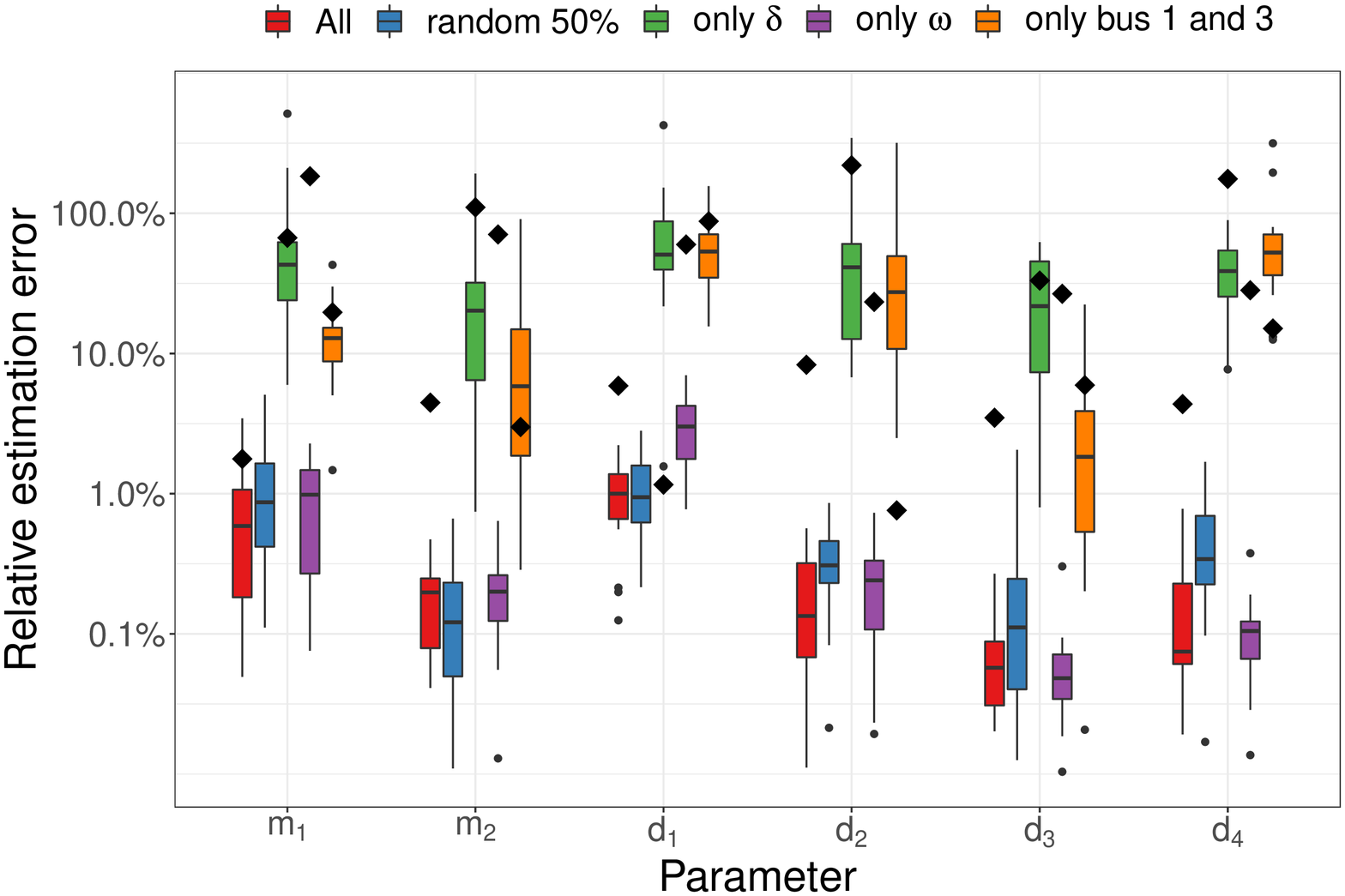}
    \caption{Estimation accuracy of PINNs (boxplots) and UKF (diamonds) with a subset of measurements. Scenario A has not been implemented for the UKF.}
    \label{fig:incomplete_data_pinn}
\end{figure}

So far, we always provided the PINN and the UKF with a complete set of measurements for $\delta_i$ and $\omega_i$ and only varied the time span and number of collocation points. Let us investigate what happens in the following scenarios where we only use:
\begin{itemize}
    \item A - (random) 50\% of the previous measurements, 
    \item B - measurements of two buses - here bus 1 and 3,
    \item C - angle measurements, or
    \item D - frequency measurements.
\end{itemize}
\Cref{fig:incomplete_data_pinn} reports the results of the various scenarios. The accuracy of the PINN remains similar for scenarios A and D compared to the full data set. In the other two scenarios the parameter estimates are mostly not as accurate, however, the large spread within each estimated parameter suggest that the initialisation and training process affect the estimates. The fact that all these scenarios do not require adjustments in the method underlines the great flexibility PINNs can offer to incorporate measurements.

\subsection{Computing time}

The flexibility and accuracy of PINNs come at the cost of a computational burden. While an efficient UKF implementation can be directly setup and executed in real time, the presented PINNs require a training time of up to tens of minutes until it yields a reliable result. This is due to the fact that a NN training is an optimisation in a highly non-convex and high-parameter space, and therefore, it is by design computationally intensive. However, the size of the network (closely related to the system size and the system's non-linearity), the number of data-points, the batching strategy and the `information content' of the measurements largely influence the required time. By tuning these parameters for each case, as for example for system A, we can train a PINN within 90 seconds.

\section{Discussion and Outlook}
Physics-informed neural networks can form the basis of a powerful additional tool for dynamic state and parameter estimation. They demonstrate higher accuracy when the system becomes increasingly non-linear, such as during low-inertia periods, are more robust against measurement noise, and perform better if there is limited data availability. However, compared to the Unscented Kalman Filtering method (UKF), PINNs are computationally expensive. Ways to address this issue are to improve the initialisation of the state estimator's weights to achieve faster convergence, and the use of Graphics Processing Units (GPUs), code optimisation and additional parallelisation. None of these has been used for the PINNs we trained in this paper.

Having investigated the performance of PINNs and UKF side-by-side in this paper, we expect that hybrid approaches, which use filter-based approaches for the online estimation but rely on PINNs (which have been trained in advance) for the system prediction can achieve the best performance, combining the best of the two methods. 

In this paper we focused on estimating the damping and inertia coefficients. However, the same methodology can be applied to estimate system parameters related to nodal voltages and line impedances, i.e. the connectivity $a_{ij}=V_i V_j B_{ij}$, given that it is known which lines are connected, i.e. $B_{ij} \neq 0$. If extended to the context of online monitoring, changes in the system parameters or the system topology could also be detected and identified. An approach in this direction, but based on a different set of methods, has already appeared in the literature \cite{Giannakis_PIDNN}.

Future work should also investigate the system theoretical aspects of this method, e.g. the observability and stability analysis. Understanding how to interpret PINNs in this context could not only contribute to improve their training procedure but also to build trust into the method which is crucial for applying it in a safety-critical context.

\section{Conclusions}
This paper introduced the application of PINNs for dynamic parameter estimation in power systems, and assessed their performance compared to other state-of-the-art dynamic state estimation and system identification methods, such as the Unscented Kalman Filter (UKF). Physics-informed neural networks combine a continuous, differentiable, and non-linear state predictor with a physics regularisation yielding a method which is highly flexible with respect to incorporating measurement data. We have shown that PINNs demonstrate a better performance during periods with stronger non-linearities, when there is limited availability of measurement data, and against measurements with high noise. Within the discussion of the results, we raised the concern of the higher computational burden, and outlined directions to mitigate it. Future work will focus on the development of hybrid approaches, combining filter-based methods for the online estimation while relying on PINNs for the system prediction, which we expect will achieve the best performance. Furthermore, we will investigate larger systems and more efficient training procedures.

\bibliographystyle{IEEEtran}
\bibliography{References}

\end{document}